\documentclass[12pt,a4paper]{article}
\usepackage{latexsym}
\usepackage{amsfonts}
\usepackage{amsmath}
\newcommand\non{\nonumber}

\newcommand{\tr}{\mbox{\rm Tr}\,}
\newtheorem{theorem}{Theorem}[section]

\def \outlineby #1#2#3{\vbox{\hrule\hbox{\vrule\kern #1%
\vbox{\kern #2 #3\kern #2}\kern #1\vrule}\hrule}}%
\def \endbox {\outlineby{4pt}{4pt}{}}%
\newcommand{\qed}{\hfill \endbox}
\newcommand{\be}{\begin{equation}}
\newcommand{\ee}{\end{equation}}
\newcommand{\bea}{\begin{eqnarray}}
\newcommand{\eea}{\end{eqnarray}}

\font\BB=msbm10

\newcommand{\PP}{\hbox{\BB P}}

\newcommand{\ZZ}{\hbox{\BB Z}}

\def\reff#1{(\ref{#1})}
\begin{document}

\begin{titlepage}

\begin{center}
{\large{\bf{Entanglement Assisted Classical Capacity of a Class }}}
\end{center}

\begin{center}
{\large{\bf{of Quantum Channels
with Long-Term Memory}}}
\end{center}
\vspace{2cm}
\bigskip
\centerline{Nilanjana Datta and Yurii Suhov} \centerline{Statistical Laboratory}
\centerline{Centre for Mathematical Sciences}
\centerline{University of Cambridge} \centerline{Wilberforce Road,
Cambridge CB30WB} \centerline{email: n.datta@statslab.cam.ac.uk}
\centerline{email: yms@statslab.cam.ac.uk}
\bigskip

\centerline{Tony C. Dorlas} \centerline{Dublin Institute for
Advanced Studies} \centerline{School of Theoretical Physics}
\centerline{10 Burlington Road, Dublin 4, Ireland.}
\centerline{email: dorlas@stp.dias.ie }
\bigskip

\vspace{1cm}

\noindent
\narrower{\bf Keywords:} quantum channels with long-term memory,
entanglement assisted classical capacity 
\bigskip

\begin{abstract} 
In this paper we evaluate the entanglement assisted classical capacity of a 
class of quantum channels with long-term memory,
which are convex combinations of memoryless channels.
The memory of such channels can be considered to be given
by a Markov chain which 
is aperiodic but not irreducible. This class of channels was
introduced in \cite{DD}, where its product state capacity was evaluated.

\end{abstract}
\end{titlepage}
%\end{document}
\newpage

\section{Introduction}

The biggest hurdle in the path of efficient information
transmission is the presence of noise, in both classical and quantum
channels. This noise causes a distortion of the information sent
through the channel. Error--correcting codes are used to overcome
this problem. Messages are encoded into codewords, which are then
sent through the channel. Information transmission is said to be
reliable if the probability of error, in decoding the output of the
channel, vanishes asymptotically in the number of uses of the channel
(see e.g. \cite{Cover} and
\cite{Nielsen}). The aim is to achieve reliable transmission,
whilst optimizing the rate, i.e., the ratio between the size of the
message and its corresponding codeword. The optimal rate of reliable
transmission is referred to as the capacity of the channel.

A classical communications channel has a unique 
capacity, the formula for which was obtained by Shannon in 1948.
A quantum channel, in contrast, has various distinct capacities. This is
because there is flexibility in the use of a quantum channel.
The particular definition of the capacity which is applicable,
depends on the following:
$(i)$ whether the information transmitted is classical or quantum; 
$(ii)$ whether the sender\footnote{We follow the normal convention and 
refer to the sender as Alice, and the receiver as Bob.}, Alice, 
is allowed to use inputs {\em{entangled}}
over various uses of the channel or whether she is only allowed to use 
product state inputs; $(iii)$ whether the receiver, Bob, is allowed to make collective 
measurements over multiple outputs of the channel or whether 
he is only allowed to measure the output of each channel use separately; $(iv)$
{whether Alice and Bob have additional resources
e.g. prior shared entanglement.

The different capacities resulting from the different choices mentioned above were
evaluated initially for memoryless\footnote{
For such a channel, the noise affecting successive input states
is assumed to be perfectly uncorrelated.} quantum channels.
The capacity of a quantum memoryless channel for transmitting classical information,
obtained under the restriction that the inputs are product states
and that collective measurements are made on the outputs, is referred to as the
product state (classical) capacity of the channel. The formula for
this capacity is given by the Holevo-Schumacher-Westmoreland (HSW) Theorem
\cite{holevo, SW97}. The formula for the quantum capacity of a memoryless channel, i.e., 
its capacity for transmitting quantum information, was established through a series 
of papers \cite{schuq, lloyd, shor, dev, hay}. The maximum asymptotic 
rate of reliable transmission of classical information with the help of 
unlimited prior entanglement between the sender and the receiver is known 
as entanglement assisted capacity. The formula for this was first 
obtained by Bennett, Shor, Smolin and Thapliyal \cite{BSST1, BSST2} and the
proof was later simplified by Holevo \cite{holea}. These proofs are based
on the HSW Theorem. For an alternative proof, based on a packing argument, 
see \cite{hsieh}.

The assumption of uncorrelated noise in quantum channels cannot be 
always justified, and memory effects should be accounted for. To our 
knowledge, the first paper concerning a quantum channel
with memory was by {Macchiavello and Palma} \cite{chiara}.
In \cite{bowen}, an important class of quantum channels with memory,
called {\em{forgetful channels}} (cf.\cite{KW}) was introduced. In such a channel,
the correlation in the noise, acting on inputs to the channel, decays
with the number of channel uses. See also \cite{DD1} and \cite{berlin}.

The capacities of channels with long-term memory (i.e., channels
which are ``not forgetful''), had remained an open problem until
recently. In \cite{DD}, the classical capacity of a class
of quantum channels with long-term memory, which 
are given by convex combinations of memoryless channels, was evaluated. 
This is perhaps the simplest class of models of ``not forgetful'' quantum
channels. For further example of such channels see \cite{DD2}.
In this paper we evaluate the entanglement-assisted classical capacity of the
same class of channels as in \cite{DD}.
For a channel $\Phi$ in this class,
$\Phi^{(n)} : {\cal B}({\cal H}^{\otimes n})
\to {\cal B}({\cal K}^{\otimes n})$ and the action of $\Phi^{(n)}$
on any state $\rho^{(n)} \in {\cal B}({\cal H}^{\otimes n})$
is given as follows:
\begin{equation} \Phi^{(n)}(\rho^{(n)}) = \sum_{i=1}^M
\gamma_i \phi_i^{\otimes n}(\rho^{(n)}),
\label{def_channel}
\end{equation}
where
$\phi_i: {\cal B}({\cal H}) \to {\cal B}({\cal K})$,
($i=1,\dots,M$) are completely positive, trace-preserving (CPT) maps 
and $\gamma_i > 0$, $ \sum_{i=1}^M
\gamma_i = 1$. Here ${\cal H}$ and ${\cal K}$ denote 
finite-dimensional Hilbert spaces and ${\cal B}({\cal H})$ denotes the algebra of 
linear operators acting on ${\cal H}$.
On using the channel, an initial random choice is made as to
which memoryless channel the successive input states are to be transmitted 
through. A
classical version of such a channel was introduced by
Jacobs \cite{jacobs} and studied further by
Ahlswede \cite{ahlswede}.

Note that the memory of the class of channels that we study, can be considered
to be given by a Markov chain which is aperiodic but not irreducible.
This can be seen as follows. Consider a quantum channel (of length $n$) with Markovian
correlated noise given by a CPT map $\Phi^{(n)} : {\cal B}({\cal H}^{\otimes n})
\to {\cal B}({\cal K}^{\otimes n})$, which is defined as follows:
$$
\Phi^{(n)}(\rho^{(n)}) = \sum_{i_1, \ldots, i_n=1}^M
q_{i_{n-1}i_{n}}\ldots q_{i_1i_2}\gamma_{i_1}
 (\phi_{i_1}\otimes ..\otimes \phi_{i_n})(\rho^{(n)}),
$$
Here $(i)$ $q_{ij}$ denote the elements of the transition matrix of a 
discrete--time Markov chain with a finite state space $I=\{1,2,\ldots,M\}$; 
$(ii)$ $\{\gamma_i\}_{i=1}^M$ denotes 
the invariant distribution of the chain, and $(iii)$
for each $i \in I$, $\phi_i :  {\cal B}({\cal H}) \to 
{\cal B}({\cal K})$ is a CPT map. Casting the channel defined by \reff{def_channel}
in this form yields
$q_{ij} = \delta_{ij}$. Hence the transition matrix of the Markov 
chain, in this case, is the identity matrix. In other words, once a particular branch $i\in\{1,\dots,M\}$ has been
chosen, the successive inputs are sent through this branch. Transition between the different branches (which correspond to the different states of the Markov Chain) is not permitted. 
The Markov chain is therefore aperiodic but 
not
irreducible. Hence the channel has long-term memory and does not lie in the class of forgetful channels.
%Clearly, such a Markov chain is not
%ergodic in the sense that it does not satisfy a limit theorem,
%i.e., $$ \sum_{i_2,\dots,i_n} q_{i_2|i_1} \dots q_{i_n|i_{n-1}}
%f(i_n) $$ does not tend to a constant for an arbitrary function
%$f$ on the state space. 

We start the main body of our paper with some preliminaries in
Section \ref{prelim}.
Our main result, giving the expression for the entanglement-assisted classical capacity
of the channels in question, is stated as a theorem in Section \ref{mainresult}.
The proofs of the converse and direct parts of this theorem our given in Sections
\ref{converse} and \ref{direct} respectively. In proving the direct part of the
theorem, we make use of the expression for the product state capacity of the 
channel, which was obtained in \cite{DD}.

\section{Preliminaries}
\label{prelim}
 The von Neumann entropy of a state $\rho$, i.e., a positive operator
of unit trace in ${\cal B}({\cal H})$, is defined as $S(\rho) :=
-\tr \rho \log \rho$, where the logarithm is taken to base $2$.
A quantum channel is given by a completely positive
trace--preserving (CPT) map $\Phi: {\cal B}({\cal H}) \to {\cal
B}({\cal K})$, where ${\cal H}$ and ${\cal K}$ are the input and
output Hilbert spaces of the channel. For any ensemble
$\{p_j,\rho_j\}$ of states $\rho_j$ chosen with probabilities
$p_j$, the Holevo $\chi$ quantity is defined as \be
\chi(\{p_j,\rho_j\}) := S\left( \sum_{j} p_j \,\rho_j
\right) - \sum_{j} p_j\, S(\rho_j). \label{Holevo} \ee 

\section{Main Result}
\label{mainresult}
As mentioned in the Introduction, in this paper we evaluate the entanglement-assisted classical
capacity of the class of channels with long-term
memory defined by \reff{def_channel}.

Consider the following protocol for the entanglement-assisted 
transmission of classical 
information through such a quantum channel. Suppose Alice and Bob share indefinitely 
many copies of an 
entangled pure state $\Psi^{AB} = |\psi^{AB}\rangle \langle \psi^{AB}| \in 
{\cal{B}}({\cal{H}}_A \otimes {\cal{H}}_B)$. Here the system $A\,(B)$, with Hilbert space
${\cal{H}}_A\,({\cal{H}}_B)$ is in Alice's (Bob's) possession and 
${\rm{dim}}\,{\cal{H}}_A= {\rm{dim}}\,{\cal{H}}_B$.  Suppose Alice has a set of
messages, labelled by the elements of the set ${\cal{M}}_n = \{1,2,
\ldots, M_n\},$ which she would like to communicate via the quantum channel
\reff{def_channel} to Bob, exploiting this shared entanglement.
For this purpose she uses  
encoding (CPT) maps ${\cal{E}}:=\{{\cal{E}}_j\}_{j=1}^J$ (where
$J$ is some positive integer) acting on ${\cal{B}}({\cal{H}}_A)$.
%chosen with probabilities $\pi_j$. 
In order to transmit her classical messages
through the quantum channel, Alice encodes each of her messages in a 
quantum state in $({\cal{H}}_A \otimes {\cal{H}}_B)^{\otimes n}$ in the 
following manner. To each $\alpha \in {\cal{M}}_n$ she assigns a quantum state
(or codeword) 
\be
\rho_\alpha^{AB;n}:= \rho_{\alpha, 1} \otimes\ldots  \otimes\rho_{\alpha, n} \in 
{\cal{B}}(( {\cal{H}}_A \otimes {\cal{H}}_B)^{\otimes n}).
\label{rho1}
\ee
%\be
%\rho_\alpha^{AB;n}:= \rho_{j_1(\alpha)}^{AB} \otimes\ldots  \otimes\rho_{j_n(\alpha)}^{AB} \in 
%{\cal{B}}(( {\cal{H}}_A \otimes {\cal{H}}_B)^{\otimes n}).
%\label{rho1}
%\ee
where
\be
\rho_{\alpha, k} = \bigl( {\cal{E}}_{j_k} \otimes id_B \bigr) \Psi^{AB},
\label{rh02}
\ee
for $k=1,\ldots, n$. Here $j_k \in \{1, \ldots, J\}$ and 
$id_B$ denotes the identity map in ${\cal{B}}({\cal{H}}_B)$.

%Thus, the probability of assigning the codeword $\rho_\alpha^{AB;n}$ 
%to the message $\alpha$ is $ \pi_{j_1(\alpha)}\ldots \pi_{j_n(\alpha)}$, 
%which generates an ensemble of quantum encodings.

Note that the codewords are states shared between Alice and Bob.
Alice then sends her part of these shared states to Bob through
$n$ subsequent uses of the quantum channel \reff{def_channel}. 
Hence, Bob's final state corresponding to Alice's classical message 
$\alpha$ is
\be
\sigma_\alpha^{AB;n} := \bigl( \Phi^{(n)} \otimes id_B^{\otimes n}\bigr)\rho_\alpha^{AB;n}.
\label{sig1}
\ee
In order to infer the message
that Alice communicated to him, Bob makes a measurement on the state $\sigma_\alpha^{AB;n}$,
the measurement being described by POVM elements $F_\alpha^{AB;n}$, $\alpha =1, \ldots, M_n$,
with $F_\alpha^{AB;n}$
being a positive operator acting in $({\cal{H}}_A \otimes {\cal{H}}_B)^{\otimes n}$, such that
$$\sum_{\alpha=1}^{M_n} F_\alpha^{AB;n} \le I_{AB}^{\otimes n},$$ 
and $I_{AB}$ denoting the identity operator acting in ${\cal{H}}_A \otimes {\cal{H}}_B$. 
Defining
$F_0^{AB;n} := (I_A \otimes I_B)^{\otimes n} - \sum_{\alpha=1}^{M_n} F_\alpha^{AB;n}$, yields a
resolution of identity in $({\cal{H}}_A \otimes {\cal{H}}_B)^{\otimes n}$. Hence,
${\cal{F}}:= \{F_\alpha^{AB; n}\}_{\alpha=0}^{M_n}$ defines a POVM. An output $\beta \in 
{\cal{M}}_n$ of a measurement described by this POVM, 
would lead Bob to conclude that the codeword was
$\rho_\beta^{AB;n}$,
whereas the output $0$ is interpreted as a failure of any
inference.

The encoding and
decoding operations, employed to achieve reliable transmission of
information by means of this protocol, together define a
quantum code ${\cal{C}}^{(n)}$ (of length $n$) which is given by the 
triple ${\cal{C}}^{(n)}:= (M_n, {\cal{E}}, {\cal{F}})$, with $M_n$ denoting
its size, and ${\cal{E}}, {\cal{F}}$ being the encoding and decoding maps employed.

Assuming equidistribution of messages, the average probability of error for the code 
${\cal{C}}^{(n)}$ is given by
\begin{equation}
P_e({\cal{C}}^{(n)}):= \frac{1}{M_n} \sum_{\alpha=1}^{M_n} \left(1 - 
{\tr}\bigl(F_\alpha^{AB;n}
(\Phi^{(n)}\otimes {id}_B^{\otimes n})(\rho_\alpha^{AB;n}) \bigr)\right).
\label{codeerr}
\end{equation}
%where ${id}_B^{\otimes n}$ denotes the identity operator in 
%${\cal{B}}({\cal{H}}_B^{\otimes n})$.
%We refer to such codes ${\cal{C}}^{(n)}$ as one-shot product-codes.
%The expected error probability
%\begin{equation}
%\overline{P_e^{(n)}} = \EE P_e({\cal{C}}^{(n)}) = \frac{1}{M_n} \sum_{\alpha=1}^{M_n} \left(1 - 
%\EE{\tr}\bigl(
%(\Phi^{(n)}_A\otimes {id}_B^{\otimes n})(\rho_\alpha^{AB;n}) F_\alpha^{AB;n}\bigr)\right),
%\end{equation}
%is obtained by further averaging over the sample codes ${\cal{C}}^{(n)}$.

If for a given $R>0$ there exists a sequence of $M_n$'s with
$$ R \le \liminf_{n \rightarrow \infty} \frac{1}{n} \log M_n,$$
and a sequence of codes ${\cal{C}}^{(n)}$ of size $M_n$ 
such that 
$$ \lim_{n \rightarrow \infty} {P_e({\cal{C}}^{(n)}})= 0,$$
then $R$ is said to be an {\em{achievable}} rate.
%under the choice of the ensemble $\{\pi_j, {\cal{E}}_j\}$ and the initial shared state $\Psi^{AB}$. 

We define the {\em{one-shot}} entanglement-assisted classical capacity \cite{holea} 
of the long-term memory channel defined by \reff{def_channel} as 
\be
C_{ea}^{(1)}(\Phi) := \sup_{\Psi^{AB}} \, \sup\bigl[ R : R \,\,{\hbox{achievable}} \bigr],
\label{oneshot}
\ee
where the internal supremum is over the rates achievable under the choice of the initial shared state $\Psi^{AB}$.\footnote{In the
case of a memoryless channel, this reduces to an alternative expression referring to a single use of the channel [cf., e.g., eq.(1) in
\cite{holea}].}

More generally Alice and Bob may share indefinitely many copies of a pure state $\Psi^{AB;m}$ in $({\cal{H}}_A \otimes {\cal{H}}_B)^{\otimes m}$
for some given $m>1$. In this case Alice can perform a similar construction 
using encoding CPT maps, ${\cal{E}}_j^{(m)}$, which act in ${\cal{H}}_A^{\otimes m}$.
In other words, she uses $m$-block encoding, and encodes a 
message $\alpha \in {\cal{M}}_n$ by the state
$$ \rho_{\alpha,m}^{AB;n}:= \rho_{\alpha,1}^{(m)} \otimes\ldots  \otimes\rho_{\alpha,n}^{(m)} \in 
{\cal{B}}(( {\cal{H}}_A \otimes {\cal{H}}_B)^{\otimes mn}),$$
where 
%$j_k(\alpha) = i$ with probability $\pi_j^{(m)}$ independently 
%for every $k=1,2,\ldots, n$,and
$$ \rho_{\alpha,k}^{(m)} = \bigl( {\cal{E}}_{j_k}^{(m)} \otimes id_B^{\otimes m} \bigr) \Psi^{AB;m},$$
for $k=1,\ldots, n$ and $j_k \in \{1, \ldots, J\}$.

As before, Bob uses decoding POVM elements $F_{\alpha,m}^{AB;n}$
which are positive operators acting in $({\cal{H}}_A{\otimes \cal{H}}_B)^{\otimes mn}$,
with $\sum_{\alpha=1}^{M_n} F_{\alpha,m}^{AB;n} \le I_{AB}^{\otimes mn}$.

The average probability of error for the resultant code (which we denote
by ${\cal{C}}_m^{(n)}$) is given by
\begin{equation}
{\overline{p}}_{e,m}^{(n)} \equiv P_e({\cal{C}}_m^{(n)}):= \frac{1}{M_n} \sum_{\alpha=1}^{M_n} \left(1 - 
{\tr}\bigl(F_{\alpha,m}^{AB;n}
(\Phi^{(mn)}\otimes {id}_B^{\otimes mn})(\rho_{\alpha,m}^{AB;n}) \bigr)\right).
\label{codeerr2}
\end{equation}

This gives rise to the $m$-{\em{shot}} entanglement-assisted classical capacity 
of the long-term memory channel defined by \reff{def_channel}:
\be
C_{ea}^{(m)}(\Phi) := \sup_{ 
\Psi^{AB;m}} \, \sup\bigl[ R : R \,\,{\hbox{achievable}} \bigr],
\label{mshot}
\ee
where the internal supremum is over the rates achievable under the choice of 
the initial shared state $\Psi^{AB;m}$. 

Finally, the full entanglement-assisted classical capacity 
of $\Phi$ is given by
\be
C_{ea}(\Phi) := {\operatornamewithlimits{\lim\sup}\limits_{{m \rightarrow 
\infty}}}\,
\frac{1}{m}\,C_{ea}^{(m)}(\Phi)  
\label{ea}
\ee
Our main result is given by the following theorem.

\begin{theorem}
\label{mainthm} The entanglement assisted classical capacity of a channel $\Phi$,
with long-term memory, defined through \reff{def_channel}, is given
by
\begin{equation}
C_{ea}(\Phi)= \max_{\rho} \left[ {\bigwedge}_{i=1}^M
I(\rho; \phi_i) \right],
\label{ans}
\end{equation}
with $I(\rho; \phi_i) := S(\rho) + S(\phi_i(\rho)) - S(\rho; \phi_i)$,
where $S(\rho; \phi_i)$ denotes the entropy exchange and is defined as follows:
\be
S(\rho; \phi_i) := S\bigl((\phi_i \otimes {id}^R) \psi^{AR}_\rho)\bigr), 
\ee
with $\psi^{AR}_\rho$ being a purification of $\rho$ on a reference system $R$.
In \reff{ans} the maximum is taken over states $\rho \in 
{\cal{B}}( {\cal{H}}_A)$.
\end{theorem} Here we use the standard notation $\bigwedge$ to
denote the minimum.  

\subsection{{Proof of the converse part of Theorem \ref{mainthm}}}
\label{converse}

In this section we prove that for any rate $R > C_{ea}(\Phi)$, with $C_{ea}(\Phi)$
given by \reff{ans}, 
reliable entanglement-assisted transmission of classical information
from Alice to Bob via the quantum channel $\Phi$ (eq.\reff{def_channel}) is impossible,
regardless of the encoding used.
 
Suppose Alice and Bob share multiple copies of an 
entangled bipartite pure state $\Psi^{AB;m}$ in  
$({\cal{H}}_A \otimes {\cal{H}}_B)^{\otimes m}$, where $m$ is a 
given positive integer. Then, given $n \in \ZZ$, Alice 
encodes her classical messages by applying chosen $m$-block encoding CPT maps, $n$ times,
to her part of the shared state $(\Psi^{AB;m})^{\otimes n}$. Here we show that the
average error probability of the corresponding code, as defined in \reff{codeerr2},
does not tend to zero as $n \rightarrow \infty$, for any $m$ and any choice
of encoding maps. For notational simplicity, we will omit 
the label $m$ and the superscript $AB$ in the rest of this section.

Let 
$$\sigma_{\alpha}^{ n}(i) := \sigma_{\alpha,1}(i) \otimes \ldots \otimes
\sigma_{\alpha,n}(i)$$
denote Bob's final state, if the 
codeword
\be
\rho_{\alpha}^{ n} = \rho_{\alpha,1}^{}\otimes \ldots\otimes \rho_{\alpha,n}^{} \,\,\in\,\,
 {\cal{B}}(({\cal{H}}_A \otimes {\cal{H}}_B)^{\otimes mn}),
\label{comp}
\ee
corresponding to the message $\alpha$, 
is transmitted through the $i$-th branch of the channel.
Here 
% $\rho_{\alpha,k} = \bigl( {\cal{E}}_k^{(m)} \otimes id_B^{\otimes m} \bigr) \Psi^{AB;m}$ 
% and 
$\sigma_{\alpha,k}(i) = (\phi_i \otimes id_B) \rho_{\alpha, k}^{}$,
for $k=1,2,\ldots , n$.
Also let 
$$\sigma_{\alpha}^{ n}:= \sum_{i=1}^M \gamma_i \sigma_{\alpha}^{ n}(i)
\quad;\quad {\bar \sigma}^{n}(i) =
\frac{1}{|{\cal M}_n|} \sum_{\alpha \in {\cal M}_n}
\sigma_{\alpha}^{n}(i)$$
$${\hbox{and}}\quad 
{\bar\sigma}_{k}(i) = \frac{1}{|{\cal M}_n|} \sum_{\alpha \in {\cal
M}_n} \sigma_{\alpha,k}(i)\ , \quad {\hbox{for }} \,\,k=1,\ldots ,n.$$

Then the average probability of error \reff{codeerr2} equals
\begin{equation}
{\bar p}_e^{(n)} := 1- \frac{1}{|{\cal M}_n|} \sum_{\alpha \in
{\cal M}_n} \tr\,\left[F_\alpha^{n} \sigma_\alpha^{n}
 \right].
\end{equation} We also define the average probability of error corresponding
to the $i^{th}$ branch of the channel as
\begin{equation}
{\bar p}_{i,e}^{(n)} := 1- \frac{1}{|{\cal M}_n|} \sum_{\alpha \in
{\cal M}_n} \tr\,\left[  \sigma_{\alpha}^{n}(i)
F_\alpha^{n} \right] \quad{\hbox{so that }}\,\, {\bar p}_e^{(n)} = \sum_{i=1}^M \gamma_i
{\bar p}_{i,e}^{(n)}
\label{errpr}
\end{equation}
%so that \begin{equation} {\bar p}_e^{(n)} = \sum_{i=1}^M \gamma_i
%{\bar p}_{i,e}^{(n)}. \label{errpr}
%\end{equation}

Let $X^{(n)}$ be a random variable with a uniform distribution
over the set ${\cal M}_n$, characterizing the classical message
sent by Alice to Bob. Let $Y_i^{(n)}$ be the random variable
corresponding to Bob's inference of Alice's message, when the
codeword is transmitted through the $i^{th}$ branch of the
channel. It is defined by the conditional probabilities
\begin{equation}
\PP\,[{Y_i^{(n)}} = \beta\,|\, X^{(n)} = \alpha] = \tr\,
[F_{\beta}^{n}\bigl(\phi_i^{\otimes n} \otimes id_B^{\otimes n}\bigr)(\rho_{\alpha}^{n}) ].
\end{equation} By Fano's inequality,
\begin{equation} h({\bar p}_{i,e}^{(n)}) + {\bar p}_{i,e}^{(n)}
\log(|{\cal M}_n|-1) \geq H(X^{(n)}\,|\, Y_i^{(n)}) = H(X^{(n)}) -
H(X^{(n)}\,:\, Y_i^{(n)}). \label{Fano}
\end{equation}
Here $h(p):= - p \log p - (1-p) \log (1-p)$ denotes the binary entropy,  
$H(A):= - \sum_a p_a \log p_a$ denotes
the Shannon entropy of a random variable $A$ with probability mass function
$p_a$, and $H(A|B)$, $H(A:B)$ denote, respectively, 
the conditional entropy and 
the mutual information \cite{Cover} of two random variables $A$ and $B$. Using the Holevo bound and the subadditivity
of the von Neumann entropy we have
\begin{eqnarray}
H(X^{(n)}\,:\,Y_i^{(n)}) &\leq& S\left(\frac{1}{|{\cal M}_n|}
\sum_{\alpha \in {\cal M}_n} \sigma_{\alpha}^{n}(i) \right) - \frac{1}{|{\cal M}_n|}
\sum_{\alpha \in {\cal M}_n} S \left(\sigma_{\alpha}^{n}(i)) \right) 
\nonumber \\ &\leq& \sum_{k=1}^n \left[ S\left({\bar \sigma}_{k}(i) \right) -
\frac{1}{|{\cal M}_n|} \sum_{\alpha \in {\cal M}_n} S
\left( \sigma_{\alpha,k}^{}(i) \right) \right] \nonumber \\
&=&  \sum_{k=1}^n \chi\left( \left\{ \frac{1}{|{\cal M}_n|},
\sigma_{\alpha,k}^{}(i) \right\}_{\alpha \in {\cal M}_n} \right) \non \\
&=& \sum_{k=1}^n \frac{1}{|{\cal M}_n|} \sum_{\alpha \in {\cal
M}_n} S \left( \sigma_{\alpha,k}^{}(i)\,||\, {\bar \sigma}^{}_{k}(i)
\right) := \sum_{k=1}^n V_k. \label{Holevosubadd}
\end{eqnarray}
In the above the symbol $S(\rho||\omega)$ denotes the quantum relative entropy 
of states $\rho$ and $\omega$.

The expression $V_k$ can be rewritten using Donald's identity \cite{don}:
\begin{equation} \sum_\alpha p_\alpha S(\omega_\alpha\,||\,\rho) = \sum_\alpha p_\alpha
S(\omega_\alpha\,||\,\bar \omega) + S({\bar \omega}\,||\, \rho),
\end{equation} where ${\bar \omega} = \sum_\alpha p_\alpha \omega_\alpha$. 
We
apply this with 
$\rho$ replaced by \begin{equation} {\bar
\sigma}^{}(i) = \frac{1}{n |{\cal M}_n|} \sum_{k=1}^n \sum_{\alpha \in
{\cal M}_n} \sigma_{\alpha,k}^{}(i), \end{equation}
$\omega_\alpha$ replaced by $ \sigma_{\alpha,k}^{}(i)$, $p_\alpha$
replaced by $1/|{\cal M}_n|$,
and consequently ${\bar \omega}$ replaced by ${\bar \sigma}^{}_{k}(i)$.
Hence, 
\begin{eqnarray}
\frac{1}{|{\cal M}_n|} \sum_{\alpha \in {\cal
M}_n} S \left( \sigma_{\alpha,k}^{}(i)\,||\, {\bar \sigma}^{}_{k}(i)
\right) &=& \frac{1}{ |{\cal M}_n|} \sum_{\alpha \in {\cal M}_n}
S(\sigma_{\alpha,k}^{}(i) \,||\, {\bar \sigma}^{}_i) - S({\bar \sigma}^{}_{k}(i)\,
||\, {\bar \sigma}^{}(i) )\nonumber\\
&\le & \frac{1}{ |{\cal M}_n|} \sum_{\alpha \in {\cal M}_n}
S(\sigma_{\alpha,k}^{}(i) \,||\, {\bar \sigma}^{}(i)),
\end{eqnarray}
where we have used the non-negativity of the quantum relative entropy.
Inserting this into \reff{Holevosubadd} we now have: 
\begin{eqnarray}
\frac{1}{n} H(X^{(n)}\,:\,Y_i^{(n)}) &\leq & \frac{1}{n |{\cal M}_n|}
\sum_{k=1}^n \sum_{\alpha \in {\cal M}_n} S(\sigma_{\alpha,k}^{}(i)
\,||\, {\bar \sigma}^{}(i))\nonumber\\
&=& \chi \left( \left\{ \frac{1}{n |{\cal
M}_n|} ,\sigma_{\alpha,k}^{}(i) \right\}_{(\alpha,k)} \right).
\end{eqnarray}
The inequality \reff{Fano} now yields (cf. eq.(17) of \cite{holea})
\begin{eqnarray} h({\bar
p}_{i,e}^{(n)}) + {\bar p}_{i,e}^{(n)} \log\,{|{\cal M}_n|} &\geq &
\log\,{|{\cal M}_n|} - n\,
\chi \left( \left\{ \frac{1}{n |{\cal
M}_n|} ,\sigma_{\alpha,k}^{}(i) \right\}_{(\alpha,k)}
\right) \nonumber\\
&\geq & \log\,{|{\cal M}_n|} - n\, I(\rho, \phi_i),
\label{holinp}
\end{eqnarray}
where $$\rho:= \sum_{\alpha, k} p_{\alpha, k} \,\rho_{\alpha, k}^A
\in {\cal{B}}({\cal{H}}_A),$$
with ${\displaystyle{p_{\alpha, k} :=  \frac{1}{n |{\cal{M}}_n|}}}$ for each $\alpha$ and $k$,
and  $\rho_{\alpha, k}^A = \tr_B \bigl(\rho_{\alpha,k}^{} \bigr)$, $k=1,\ldots, n$.
 However, since
\begin{equation} C_{ea}(\Phi) \geq \bigwedge_{i=1}^M I(\rho, \phi_i)
\end{equation} and $R = \frac{1}{n} \log |{\cal M}_n| > C_{ea}(\Phi)$,
there must be at least one branch $i$ such that
\begin{equation} {\bar p}_{i,e}^{(n)} \geq 1 - \frac{C_{ea}(\Phi)+
{1}/{n}}{R} > 0.\label{errpr2} \end{equation} We conclude from
(\ref{errpr}) and \reff{errpr2} that \begin{equation} {\bar
p}_e^{(n)} \geq  \left(1 - \frac{C_{ea}(\Phi)+ {1}/{n}}{R}
\right)\,\bigwedge_{i=1}^M {\gamma_i}.
\end{equation}
Hence ${\bar p}_{i,e}^{(n)}$ does not tend to zero as $n \rightarrow \infty$, which
in turn implies that
$$C_{ea}(\Phi) \le  \max_{\rho} \left[ {\bigwedge}_{i=1}^M
I(\rho; \phi_i) \right].$$

\qed
\subsection{Proof of the direct part of Theorem \ref{mainthm}}
\label{direct}
In this section we prove that $C_{ea}(\Phi)$, defined by \reff{ea}
satisfies the lower bound
\be
 C_{ea}(\Phi) \ge  \max_{\rho} \left[ {\bigwedge}_{i=1}^M
I(\rho; \phi_i) \right],
\ee
where the maximum is taken over all states $\rho \in {\cal{B}}({\cal{H}}_A)$.

To prove this we employ the following result which we proved in \cite{DD}:
\begin{theorem}
\label{mainthm2} The product state capacity of a channel $\Phi$,
with long-term memory, defined through \reff{def_channel}, is given
by
\be
C(\Phi)= \sup_{\{\pi_j,\rho_j\}} \left[ {\bigwedge}_{i=1}^M
\chi_i(\{\pi_j,\rho_j\}) \right],
\label{ddresult}
\ee
where $\chi_i(\{\pi_j,\rho_j\}):= \chi\left(\{\pi_j, \phi_i(\rho_j)\}\right)$.
The supremum is taken
over all finite ensembles of states $\rho_j\in {\cal{B}}({\cal{H}})$, chosen  
with probabilities
$\pi_j$.
\end{theorem} 
%Here the notation $\bigwedge$ denotes the minimum.

From the definition \reff{oneshot} of the one-shot entanglement assisted capacity
and \reff{ddresult} it follows that
\be
C_{ea}^{(1)}(\Phi)= \sup_{\{\pi_j,{\cal{E}}_j\}, \Psi^{AB}} \left[ {\bigwedge}_{i=1}^M
\chi(\{\pi_j,\bigl(\phi_i \otimes id_B\bigr)\rho_j^{AB}\}) \right],
\ee
where (i) $\Psi^{AB}$ is the bipartite entangled pure state, indefinitely many copies
of which are shared by Alice and Bob, and (ii) ${\cal{E}}_j$ are encoding maps 
acting on ${\cal{B}}({\cal{H}}_A)$, as described in Section \ref{mainresult}, i.e.,
$\rho_j^{AB} = ({\cal{E}}_j \otimes id_B)\Psi^{AB}$.

Moreover, from the definition \reff{mshot} of the $m$-shot entanglement assisted capacity
it follows that
\be
C_{ea}^{(m)}(\Phi)= \sup_{\{\pi_j^{(m)},{\cal{E}}_j^{(m)}\}, \Psi^{AB; m}} \left[ {\bigwedge}_{i=1}^M
\chi(\{\pi_j^{(m)},\bigl(\phi_i^{\otimes m} \otimes id_B^{\otimes m}\bigr)\rho_j^{AB,m}\}) \right].
\ee
Now, following \cite{BSST2} and \cite{holea}, consider a specific encoding ensemble 
$\{\pi_{(a,b)}^{(m)}, {\cal{E}}^{(m)}_{(a,b)}\}$,
where $a,b=1,2,  \ldots, q$, for some integer $q$, and 
$$\pi_{(a,b)}^{(m)} = \frac{1}{q^2}\quad ; \quad {\cal{E}}_{(a,b)}^{(m)} = W_{a,b}^{(m)}.$$
Here $W_{a,b}^{(m)}$ denotes the discrete Weyl-Segal operators (see e.g.\cite{holea}) for 
a $q$-dimensional subspace ${\cal{Q}}_m$ of
${\cal{H}}_A^{\otimes m}$.
Further, consider the codewords to be given by
$$\varrho^{AB,m}_{a,b} = \bigl( {W}_{a,b}^{(m)} \otimes id_B^{\otimes m}\bigr)(|\psi_m^{AB}\rangle
\langle \psi_m^{AB}|),
$$
where $|\psi^{AB}_m\rangle$ denotes a maximally
entangled state of rank $q$:
$$|\psi^{AB}_m\rangle := \frac{1}{\sqrt{q}} \sum_{k=1}^q 
|e_k^{(m)}\rangle \otimes |e_k^{(m)}\rangle,
$$
where $\{|e_k^{(m)}\rangle\}_{k=1}^q$ is an orthonormal system of vectors
in ${\cal{Q}}_m$.
Hence, 
\be
C_{ea}^{(m)}(\Phi) \ge {\bigwedge}_{i=1}^M
\chi(\{\frac{1}{q^2},\bigl(\phi_i^{\otimes m} \otimes id_B^{\otimes m}\bigr)\varrho_{a,b}^{AB,m}\})
\label{eq1}
\ee
From \cite{holea} it follows that
\be
\chi(\{\frac{1}{q^2},\bigl(\phi_i^{\otimes m} \otimes id_B^{\otimes m}\bigr)\varrho_{a,b}^{AB,m}\})
= I\bigl(\frac{P^{(m)}}{\tr(P^{(m)})}; \phi_i^{\otimes m}\bigr),
\label{eq2}
\ee
where $P^{(m)}$ is the orthoprojection onto ${\cal{Q}}_m$. Further, it was proved in \cite{holea}
that if ${\cal{Q}}_m$ is chosen to be the strongly $\delta$-typical subspace for an 
arbitrary state $\rho^{\otimes m} \in {\cal{B}}({\cal{H}}_A^{\otimes m})$, and ${P^{m, \delta}}$
is its orthoprojection, then
\be
\lim_{\delta \rightarrow 0} \lim_{m \rightarrow \infty} \frac{1}{m} 
I\bigl(\frac{P^{m, \delta}}{\tr(P^{m,\delta})}; \phi_i^{\otimes m}\bigr)
= I(\rho; \phi_i)
\label{eq3}
\ee 
From \reff{eq1}, \reff{eq2}, \reff{eq3} and the definition \reff{ea} of the full entanglement-assisted 
capacity, it follows that
\be
C_{ea}(\Phi) \ge  {\bigwedge}_{i=1}^M I(\rho; \phi_i).
\ee
\qed
\section*{Acknowledgements} The authors would like to thank Alexander Holevo
for helpful comments. YMS thanks the Isaac Newton Institute, University of Cambridge, for hospitality
during the Spring of 2007.

\end{document}